\documentclass{emulateapj}
\usepackage{multirow}
\usepackage{subfigure}
\def\msun{$M_{\odot}$}

\def\ergsec{\hbox{erg s$^{-1}$}} 
\def\ergscm{\hbox{erg s$^{-1}$ cm$^{-2}$}} 
\def\ergscma{\hbox{erg s$^{-1}$ cm$^{-2}$ \AA$^{-1}$}}

\def\xmm{{\it XMM-Newton}}
\newcommand\nar{\ref@jnl{New A Rev.}}%
\tabletypesize{\tiny}
\shortauthors{Lin et al.}
\begin{document}

\title{A $\sim$3.8 hour Periodicity from an Ultrasoft Active Galactic Nucleus Candidate}

\author{Dacheng Lin\altaffilmark{1}, Jimmy A. Irwin\altaffilmark{1}, Olivier Godet\altaffilmark{2,3}, Natalie A. Webb\altaffilmark{2,3}, Didier Barret\altaffilmark{2,3}}
\altaffiltext{1}{Department of Physics and Astronomy, University of Alabama, Box 870324, Tuscaloosa, AL 35487, USA, email: dlin@ua.edu}
\altaffiltext{2}{CNRS, IRAP, 9 avenue du Colonel Roche, BP 44346, F-31028 Toulouse Cedex 4, France}
\altaffiltext{3}{Universit\'{e} de Toulouse, UPS-OMP, IRAP, Toulouse, France}

\begin{abstract}
Very few galactic nuclei are found to show significant X-ray quasi-periodic oscillations (QPOs). After carefully modeling the noise continuum, we find that the $\sim$3.8 hr QPO in the ultrasoft active galactic nucleus (AGN) candidate \object{2XMM J123103.2+110648} was significantly detected ($\sim$$5\sigma$) in two \textit{XMM-Newton} observations in 2005, but not in the one in 2003. The QPO rms is very high and increases from $\sim$25\% in 0.2--0.5 keV to $\sim$50\% in 1--2 keV. The QPO probably corresponds to the low-frequency type in Galactic black hole X-ray binaries, considering its large rms and the probably low mass ($\sim$$10^5$ \msun) of the black hole in the nucleus. We also fit the soft X-ray spectra from the three \textit{XMM-Newton} observations and find that they can be described with either pure thermal disk emission or optically thick low-temperature Comptonization. We see no clear X-ray emission from the two \textit{Swift} observations in 2013, indicating lower source fluxes than those in \textit{XMM-Newton} observations.

\end{abstract}


\section{INTRODUCTION}
\label{sec:intro}

Although a large variety of X-ray quasi-periodic oscillations (QPOs)
have been observed in Galactic black hole X-ray binaries (BHBs), they
are hardly seen in galactic nuclei, which are believed to harbor
supermassive black holes (SMBHs, black hole (BH) mass
$M_\mathrm{BH}\gtrsim10^5$ \msun). If BH-mass scaling works, QPOs from
SMBHs will have much longer timescales and thus be much better
resolved in time than those from BHBs. Therefore QPOs from SMBHs can
shed new lights on the origin of QPOs and in turn on the behavior of
accretion flows onto BHs. Currently the two most confident cases are
the $\sim$1.0 hr QPO from the active galactic nucleus (AGN) \object{RE
  J1034+396} \citep[$\sim$$5.6\sigma$,][but see
  \citet{va2010}]{gimiwa2008} and the $\sim$200 s one from
\object{Swift J164449.3+573451}
\citep[$\sim$$4.3\sigma$,][]{remire2012}, a tidal disruption event
(TDE) candidate, in which a star approaching the central SMBH was
tidally disrupted and subsequently accreted.

We discovered \object{2XMM J123103.2+110648} (J1231+1106 hereafter) as
a very soft source, untypical of AGNs, in our project of
classifications of 4330 X-ray sources from the 2XMMi-DR3 catalog
\citep{liweba2012}. We have reported some results of our study of this
source in \citet{liweba2013b}, including the discovery of a possibly
strong but transient QPO and very soft spectra (characteristic
blackbody (BB) temperatures of 0.1--0.15 keV, no significant emission
above 2 keV) from three \xmm\ observations. This source was also
independently discovered by \citet{tekaaw2012}, who also reported its
very soft spectra and possible presence of a QPO, which, however, did
not seem statistically significant to them. They suggested the source
as an AGN, considering its coincidence with a slightly extended SDSS
source. \citet{hokite2012} obtained a Magellan optical spectrum of
this counterpart in 2012, which exhibited as a Type 2 AGN (redshift
$z=0.11871$, the source luminosity distance of 532 Mpc, assuming a
flat universe with $H_0$=73 km~s$^{-1}$~Mpc$^{-1}$ and
$\Omega_\mathrm{M}$=0.27). The narrow lines have very small velocity
dispersions ($\sigma=33.5$ km~s$^{-1}$ for [OIII] $\lambda$5007),
suggesting a small BH mass ($\sim10^5$ \msun).

In this Letter we continue to study J1231+1106. Different from
\citet{liweba2013b}, we calculate the significance of the QPO
formally, carry out detailed spectral fits, concentrating on the
physical model by \citet{dodaji2012}, and present two \textit{Swift}
follow-up observations. In Section~\ref{sec:reduction}, we describe
the data analysis. In Section~\ref{sec:res}, we present the
results. The conclusions and the discussion of the source nature are
given in Section~\ref{sec:conclusion}.
\section{DATA ANALYSIS}
\label{sec:reduction}

\begin{deluxetable}{lccc}
\tablecaption{Properties of J1231+1106 in three \xmm\ observations\label{tbl:sppds107102}}
\tablewidth{0pt}
\tablehead{ & \colhead{0145800101} &\colhead{0306630101}&\colhead{0306630201}\\
& \colhead{(XMM1)} & \colhead{(XMM2)} & \colhead{(XMM3)}}
\startdata
 Observation Date & 2003-07-13 & 2005-12-13 & 2005-12-17 \\
 Exposure (ks, pn/MOS1/MOS2) & 45.4/58.0/61.4 & 54.8/-/68.7 & 80.8/-/92.2 \\
\hline
\\
Power Spectra: \\
\hline
bin size ($\mu$Hz) & 18.2 & 14.6 & 10.4 \\
QPO quality factor  & \nodata & $>5$ & $>7$ \\
hard lag (ks)\tablenotemark{a} & \nodata & 0.4$\pm$0.6 & 1.1$\pm$0.5 \\
QPO rms (\%, 0.2--2 keV)\tablenotemark{b} & $0\pm9$ & $30.8\pm0.5$ & $24.7\pm0.3$ \\
QPO rms (\%, 0.2--0.5 keV)\tablenotemark{b} & $0\pm9$ & $27.8\pm0.4$ & $24.2\pm0.4$\\
QPO rms (\%, 0.5--1 keV)\tablenotemark{b} & $0\pm12$  & $32.1\pm0.5$ & $21.8\pm0.8$ \\
QPO rms (\%, 1--2 keV)\tablenotemark{b} & $0\pm20$ &  $44\pm2$ & $56\pm2$\\
\hline
\\
\multicolumn{4}{l}{Energy Spectral fits\tablenotemark{c}:}\\
\hline
\multicolumn{4}{l}{Model (a): BB}\\
$N_\mathrm{H}$ (10$^{20}$ cm$^{-2}$) & \multicolumn{3}{c}{$<0.4$}\\
$kT_\mathrm{BB}$ (keV) & $0.125\pm0.005$  & $0.151\pm0.004$  & $0.134\pm0.003$ \\
$N_\mathrm{BB}$  & $42\pm10$  & $37\pm5$  & $47\pm6$ \\
$\chi^2_\nu(\nu)$ &$1.29( 72)$&$1.16(142)$&$1.08(144)$\\
\hline
\multicolumn{4}{l}{Model (b): MCD} \\
$N_\mathrm{H}$ (10$^{20}$ cm$^{-2}$) & \multicolumn{3}{c}{$0.6^{+1.8}$}\\
$kT_\mathrm{MCD}$ (keV) & $ 0.16\pm0.01$  & $ 0.20\pm0.01$  & $ 0.18\pm0.01$\\
$N_\mathrm{MCD}$  & $12^{+7}_{-3}$  & $9^{+4}_{-2}$  & $13^{+6}_{-3}$ \\
$\chi^2_\nu(\nu)$ &$1.15( 72)$&$1.04(142)$&$1.04(144)$ \\
$F_\mathrm{abs}$ (10$^{-13}$ \ergscm)\tablenotemark{d} & $ 0.48\pm0.03$  & $ 1.06\pm0.04$  & $ 0.77\pm0.03$ \\
$F_\mathrm{unabs}$ (10$^{-13}$ \ergscm)\tablenotemark{d} & $ 0.65^{+ 0.10}_{-0.06}$  & $ 1.38^{+ 0.18}_{-0.09}$  & $ 1.04^{+ 0.15}_{-0.08}$ \\
\hline
\multicolumn{4}{l}{Model (c): optxagnf (non-rotating, pure thermal disk)} \\
$N_\mathrm{H}$ (10$^{20}$ cm$^{-2}$) & \multicolumn{3}{c}{$0.2^{+1.8}$}\\
$M_\mathrm{BH}$ (\msun) & \multicolumn{3}{c}{$(4.2^{+0.9}_{-0.2})\times10^4$}\\
$L_\mathrm{Bol}/L_\mathrm{Edd}$ & $0.92\pm0.05$  & $1.77\pm0.08$  & $1.3\pm0.05$\\
$\chi^2_\nu(\nu)$ &$1.16( 72)$&$1.09(142)$&$1.07(144)$\\
\hline
\multicolumn{4}{l}{Model (d): optxagnf (maximally-rotating,  pure thermal disk)} \\
$N_\mathrm{H}$ (10$^{20}$ cm$^{-2}$) & \multicolumn{3}{c}{$0.2^{+1.8}$}\\
$M_\mathrm{BH}$ (\msun) & \multicolumn{3}{c}{$(3.2^{+0.7}_{-0.2})\times10^5$}\\
$L_\mathrm{bol}/L_\mathrm{Edd}$ & $0.110\pm0.006$ & $0.213\pm0.009$  & $0.162\pm0.007$\\
$\chi^2_\nu(\nu)$ & $1.16( 72)$&$1.09(142)$&$1.07(144)$ \\
\hline
\multicolumn{4}{l}{Model (e): optxagnf (non-rotating, $r_\mathrm{cor}=50r_\mathrm{g}$, $M_\mathrm{BH}=10^5$ \msun)} \\
$N_\mathrm{H}$ (10$^{20}$ cm$^{-2}$) & \multicolumn{3}{c}{$2^{+5}$}\\
$L_\mathrm{bol}/L_\mathrm{Edd}$ & $0.6^{+0.4}_{-0.2}$  & $1.0^{+0.6}_{-0.3}$ & $0.8^{+0.5}_{-0.2}$ \\
$kT_\mathrm{e}$ (keV)  & $ 0.14\pm0.01$ & $ 0.18\pm0.01$ & $ 0.16\pm0.01$ \\
$\tau$ & \multicolumn{3}{c}{$29\pm10$} \\
$\chi^2_\nu(\nu)$ &$1.17( 71)$&$1.05(141)$&$1.05(143)$\\
\hline
\multicolumn{4}{l}{Model (f): optxagnf (non-rotating, $r_\mathrm{cor}=50r_\mathrm{g}$, including Sw2/UVW2)} \\
$N_\mathrm{H}$ (10$^{20}$ cm$^{-2}$) & \multicolumn{3}{c}{$2^{+5}$}\\
$M_\mathrm{BH}$ (\msun) & \multicolumn{3}{c}{$(2.0^{+1.9}_{-1.1})\times10^6$}\\ 
$L_\mathrm{bol}/L_\mathrm{Edd}$ & $0.07^{+0.15}_{-0.04}$ & $0.10^{+0.19}_{-0.07}$ & $0.09^{+0.20}_{-0.05}$ \\
$kT_\mathrm{e}$ (keV) & $0.15\pm0.01$  & $0.18\pm0.01$  & $0.16\pm0.01$ \\
$\tau$ & \multicolumn{3}{c}{$25\pm8$} \\
$\chi^2_\nu(\nu)$ &$1.19( 70)$&$1.05(140)$&$1.06(142)$\\
\hline
\\
\multicolumn{4}{l}{\textit{XMM-Newton} OM Photometry:} \\
\hline
\multirow{2}{*}{V flux\tablenotemark{e} \& AB Mag} & $2.2\pm1.3$, & $3.8\pm1.1$, & $3.0\pm1.1$, \\
 & $20.5\pm0.6$ & $20.0\pm0.3$ & $20.2\pm0.4$\\
\hline
\multirow{2}{*}{B flux\tablenotemark{e} \& AB Mag} & \nodata & $2.2\pm0.6$, & $2.0\pm0.7$, \\
 &  \nodata & $21.0\pm0.3$ & $21.1\pm0.4$ \\
\hline
\multirow{2}{*}{U flux\tablenotemark{e} \& AB Mag} & $1.4\pm0.7$, & $1.2\pm0.6$, & $2.0\pm0.6$,\\
 & $22.0\pm0.6$ & $22.2\pm0.6$ & $21.6\pm0.3$ \\
\hline
\multirow{2}{*}{UVW1 flux\tablenotemark{e} \& AB Mag} & $1.6\pm1.0$, & \nodata & $2.1\pm0.9$,\\
 & $22.2\pm0.7$ & \nodata & $22.0\pm0.5$
\enddata 
\tablecomments{\textsuperscript{a}The time lag of hard X-rays (1--2 keV) behind soft X-rays (0.2--0.5 keV), with 1$\sigma$ errors from the red and Poisson noises. \textsuperscript{b}The QPO fractional rms (Poisson and red noises subtracted) using the power at the bin containing the QPO centroid frequency, with $1\sigma$ errors from the red and Poisson noises. \textsuperscript{c}All errors are at a 90\% confidence level. \textsuperscript{d}The 0.3-10 keV absorbed ($F_\mathrm{abs}$) and unabsorbed ($F_\mathrm{unabs}$) fluxes. \textsuperscript{e}In units of $10^{-17}$ \ergscma.}
\end{deluxetable}

\begin{figure*} 
\centering
\includegraphics[width=5.0in]{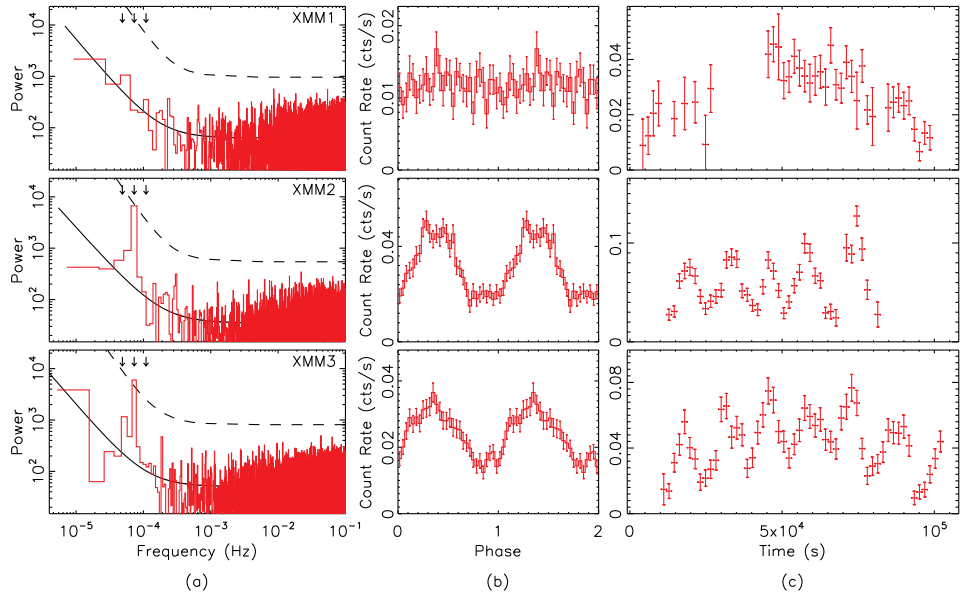}
\caption{(a): the power spectra normalized to (rms/mean)$^2$/Hz. The arrows mark the periods of $3P_0/2$, $P_0$, $2P_0/3$ ($P_0$=13710~s, see Figure~\ref{fig:107102foldcurve2}a). The solid lines are the best-fitting PL model to the red noise plus Poisson noise. The dashed lines are the 99.9\% confidence detection level. The power spectrum of XMM1 is from the second segment of data (after 40 ks in the light curve plot in the top right panel) because of frequent strong background flares in earlier data. (b): the light curves folded at $P_0$. (c): the unfolded light curves, shifted in time to be in phase. All panels use 0.2-2 keV events from all available cameras, except (c), which uses only the pn data. \label{fig:107102foldcurve}}
\end{figure*}

\begin{figure*} 
\centering
\includegraphics[width=5.0in]{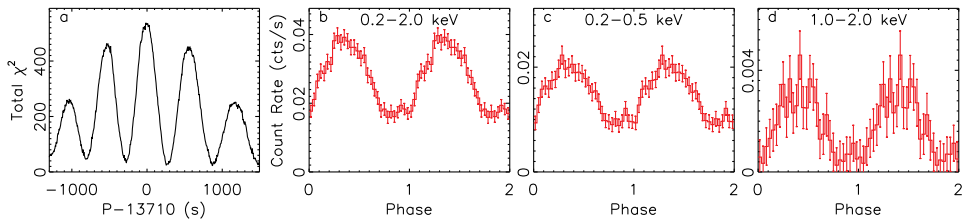}
\caption{(a): the total $\chi^2$ values from the fits with a constant to the 0.2--2 keV light curves folded at various tentative periods. (b)--(d): the 0.2--2 keV, 0.2--0.5 keV and 1--2 keV light curves folded at $P_0$=13710~s, respectively. All panels use XMM2 and XMM3 data only. \label{fig:107102foldcurve2}}
\end{figure*}

J1231+1106 was serendipitously detected at off-axis angles of
$\sim$6$\arcmin$ in three \xmm\ observations (XMM1, XMM2, and XMM3
hereafter; see Table~\ref{tbl:sppds107102}) of the quasar \object{LBQS
  1228+1116}. XMM1 was made in 2003 July, and XMM2 and XMM3 in 2005
December, with only four days apart. We used the X-ray light curves,
energy and power spectra obtained in \citet{liweba2013b}, which we
refer to for details. One exception is that here we combined data from
all available cameras (i.e., pn, MOS1 and MOS2; the source was not in
the field of view of MOS1 in XMM2 and XMM3) and used the MOS time
resolution (2.6 s) as the light curve bin size to produce the power
spectra.

The QPO in J1231+1106 has most power in one frequency bin in the
unbinned power spectra and has some underlying red noise, except the
Poisson noise. To quantify its significance, we used the maximum
likelihood method \citep{va2005,bava2012}, in which, one first obtains
the probability distribution of the true power continuum through
maximum likelihood fitting of the power spectra and then calculates
the QPO significance assuming that the observed power, after being
normalized by the true power and multiplied by a factor of 2, follows
a $\chi^2$ distribution with two degrees of freedom. We fitted the
spectra below 0.1 Hz, ignoring five frequency bins centering on the
QPO centroid frequency in the fits, in order to exclude the possible
QPO contamination. We focused on a single powerlaw (PL) model
(depending on the frequency $f$ in the form of
$f^{-\Gamma_\mathrm{PL}}$) for the red noise. The best-fitting values
of $\Gamma_\mathrm{PL}$ are $\sim$1.5 but not well constrained. In the
final fits we assumed $\Gamma_\mathrm{PL}$ to be within 1--2, a range
often seen in AGNs \citep[e.g.,][]{gimiwa2008, mc2010}. To assess the
limits of our modeling of the red noise, we also tested a PL model
with $0<\Gamma_\mathrm{PL}<3$ and a broken PL model with the indices
below ($\Gamma_\mathrm{1,PL}$) and above ($\Gamma_\mathrm{2,PL}$) the
break frequency assuming typical values (($\Gamma_\mathrm{1,PL}$,
$\Gamma_\mathrm{2,PL}$)=(1, 2), (1, 3), and (0, 2)) seen in the
literature \citep{gimiwa2008,mc2010,remire2012}. The break frequency
was allowed to be free. The Poisson noise was modeled with a constant
fixed at the value inferred from power above 0.1 Hz (it deviated from
the expected value by $<$0.3\%). To further account for the effect of
data incompleteness due to background flares, we carried out Monte
Carlo simulations following \citet{tiko1995}. For each model each
observation, $10^7$ light curves following the power spectral
distribution inferred above from the maximum likelihood method and
having the same mean count rate, variance, and sampling pattern as the
observed light curves were generated and then used to produce the
power spectra and obtain the QPO significance based on the
distribution of simulated power at the QPO frequency.

We fitted the X-ray spectra with several models: a single-temperature
BB, a multicolor disk (MCD), and the AGN spectral model optxagnf by
\citet{dodaji2012}. They are \textit{bbodyrad}, \textit{diskbb}, and {\it
  optxagnf}, respectively, in XSPEC. Both the BB and MCD models were
redshifted by $z=0.11871$ using the \textit{zashift} model in XSPEC (the
optxagnf model has handled this internally). All models included the
Galactic absorption fixed at $N_\mathrm{H}=2.3\times10^{20}$ cm$^{-2}$
\citep{kabuha2005} using the \textit{wabs} model. Possible absorption
intrinsic to the source was accounted for using the \textit{zwabs} model,
with the column densities tied to be the same in all observations
because they were consistent within the 90\% errors.

We also obtained the magnitudes and fluxes in the Optical Monitor
\citep[OM;][]{mabrmu2001} filters, with the \textit{omichain} task in the
SAS 11.0.0 package (Table~\ref{tbl:sppds107102}). We note that the B
and UVW1 filters were not used in XMM1 and XMM2, respectively, and
that the source was not detected in the UVM2 or UVW2 filters in any
observation.

At our request, \textit{Swift} \citep{gechgi2004} observed J1231+1106
twice, once on 2013 March 8 (observation ID: 00032732001, Sw1
hereafter) and the other on 2013 June 21 (observation ID: 00032732001,
Sw2 hereafter). The X-ray telescope \citep[XRT;][]{buhino2005} was
operated in Photon Counting mode for 8.5 ks and 4.4 ks,
respectively. The UV-Optical Telescope \citep[UVOT;][]{rokema2005}
used the UVW1 filter (8.3 ks) in Sw1 and the UVW2 filter (4.3 ks) in
Sw2. We analyzed the data with FTOOLS 6.13 and the calibration files
of 2013 July. The source was hardly detected in the XRT. We calculated
the count rate confidence intervals using Bayesian statistics
\citep{krbuno1991}, with radii of 20$\arcsec$ and 2$\arcmin$ for the
circular source and background regions, respectively. The UV
magnitudes and fluxes were measured with the task \textit{uvotsource}
with radii of 5$\arcsec$ and 25$\arcsec$ for the circular source and
background regions, respectively.

\section{RESULTS}
\label{sec:res}

\subsection{The $\sim$3.8 hr QPO}
Figures~\ref{fig:107102foldcurve} and \ref{fig:107102foldcurve2} show
the timing properties of J1231+1106 \citep[see also Figure 6
  in][]{liweba2013b}. As obtained by \citet{liweba2013b}, the source
showed a large coherent oscillation at a period about 3.8 hr in the
two observations (XMM2 and XMM3) in 2005 December but not clearly in
the one (XMM1) in 2003 July. The QPO concentrates in one frequency bin
(Figure~\ref{fig:107102foldcurve}a), and the quality factor, defined
as the centroid frequency divided by the frequency width (full-width
at half-maximum), is $>7$ (Table~\ref{tbl:sppds107102}). The
oscillation is only quasi-periodic, considering that the minima and
maxima of the light curves in XMM2 and XMM3 do not seem to be well in
phase (Figure~\ref{fig:107102foldcurve}c).

Assuming blind search over frequencies below 0.1 Hz (the highest
stable orbital frequency around a non-rotating BH of $2\times10^4$
\msun) in each observation separately, we obtained the 99.9\%
confidence limit in Figure~\ref{fig:107102foldcurve}a (the dashed
line) and the QPO significance at the 3.1$\sigma$ (Gaussian
probability, the same below) and 4.0$\sigma$ levels in XMM2 and XMM3,
respectively, assuming the PL model for the red noise
($1<\Gamma_\mathrm{PL}<2$, 0.2--2 keV). Considering that the QPO was
detected at the same frequency in two observations, its global
significance is 6.7$\sigma$. The Monte Carlo simulations confirmed the
significance to be $>$6.2$\sigma$. The significance is 5.1$\sigma$ if
we assumed $0<\Gamma_\mathrm{PL}<3$. Using the broken PL model for the
red noise, the QPO significance is 6.4$\sigma$, 5.4$\sigma$ and
5.9$\sigma$ for ($\Gamma_\mathrm{1,PL}$, $\Gamma_\mathrm{2,PL}$) = (1,
2), (1, 3) and (0, 2), respectively, from Monte Carlo simulations. The
fractional root-mean-square (rms) variability in the QPO is remarkably
high and increases from $\sim$25\% in 0.2--0.5 keV to $\sim$50\% in
1--2 keV in both XMM2 and XMM3 but is consistent with zero in XMM1
(Table~\ref{tbl:sppds107102}). The folded light curves in these two
energy bands are plotted in Figure~\ref{fig:107102foldcurve2}, and we
see no clear time lag between them ($\lesssim$1.1 ks from
Table~\ref{tbl:sppds107102}).

\begin{figure*}
\centering
\includegraphics[width=0.8\textwidth]{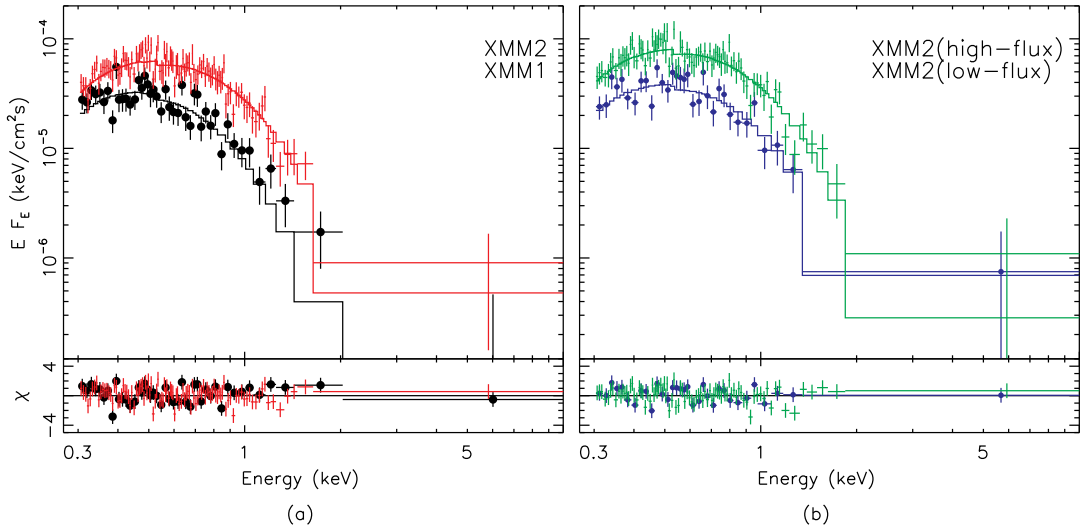}
\caption{The unfolded spectra and the fit residuals of J1231+1106 using the MCD model. (a): XMM1 (black filled circles) and XMM2 (red); (b) the spectrum in the high-flux interval (green) and that in the low-flux interval (blue filled circled) in XMM2. For clarity, we show the pn camera only and have the data rebinned. \label{fig:107102spfits}}
\end{figure*}

\subsection{The Ultrasoft X-ray Spectra}
\label{sec:spfit}
The fitting results of the \xmm\ observations with various models are
given in Table~\ref{tbl:sppds107102}. We concentrated on X-ray
spectra, because in optical the source appeared red
(Table~\ref{tbl:sppds107102}) and should be dominated by galaxy
emission, instead of nuclear accretion, as often seen in low-mass AGNs
\citep{dodaji2012}. In the UV, the source was detected in the UVW1
filter, but with large uncertainties. It still appeared a little red
in the UVW1 and UVW2 filters from the two \textit{Swift} observations
(Section~\ref{sec:swobs}). Thus we will refer to the
shortest-wavelength detection, i.e., the UVW2 one in Sw2, as possible
emission from nuclear accretion.

All three spectra are very soft, with $kT_\mathrm{BB}\sim$ 0.13--0.15 keV
or $kT_\mathrm{MCD}\sim$ 0.16--0.20 keV (source rest frame). The fits
with the BB model show systematic residuals, which are unseen using
the MCD model (Figure~\ref{fig:107102spfits}a). Both models infer
little intrinsic absorption. We see that the inner disk temperature
varies at a significance level of $4.8\sigma$ while the inner disk
radius is consistent within the 90\% confidence errors in these
observations, implying the disk luminosity $L$ to approximately follow
the relation of $L\propto T_\mathrm{MCD}^4$. The source reached a peak
0.3--10 keV absorbed luminosity of 3.6$\times$10$^{42}$ \ergsec\ and a
bolometric unabsorbed luminosity of 8.5$\times$10$^{42}$ \ergsec\ (the
MCD model) in XMM2. XMM1 is the faintest, with a 0.3--10 keV absorbed
luminosity of 1.6$\times$10$^{42}$ \ergsec\ and a bolometric unabsorbed
luminosity of 5.0$\times$10$^{42}$ \ergsec. Both the BB and MCD models
under-predicted the Sw2/UVW2 flux, by a factor of $\sim$$2\times10^4$
and 48, respectively (using the best-fitting model to XMM1 and
assuming the reddening $\mathrm{E(B-V)}=1.7\times 10^{-22} N_\mathrm{H}$,
the same below).

We used the MCD model to check the spectral variability in the
oscillation. We created and fitted two spectra from XMM2 from the
high-flux and low-flux intervals, corresponding to the pn 1 ks 0.2--2
keV count rate higher and lower than 0.05 counts~s$^{-1}$,
respectively. The column density was fixed at the value obtained above
from the simultaneous fit to all three \xmm\ observations. The
best-fitting models are shown in Figure~\ref{fig:107102spfits}b. We
inferred a lower temperature and a smaller inner radius of the disk in
the low-flux interval than in the high-flux interval, but only at a
1.7$\sigma$ confidence level.

We next look at the optxagnf model. It assumes that the gravitational
energy released in the disk is emitted as a color-corrected BB down to
a (coronal) radius $r_\mathrm{cor}$, while within this radius the
available energy is distributed between powering two Comptonization
components: the soft one via Comptonization in an optically thick cool
corona and the hard one in an optically thin hot corona.

Considering the very soft spectra, we first investigated the scenario
of pure thermal disk emission (setting $r_\mathrm{cor}$ to be at the
innermost stable circular orbit (ISCO)). The fitting results for two
cases are given in Table~\ref{tbl:sppds107102}: a non-rotating
Schwarzschild BH (Models (c)) and a maximally-rotating Kerr BH (Models
(d)). They inferred $M_\mathrm{BH}=4\times10^4$ \msun\ and $3\times10^5$
\msun, and the Eddington ratio $L_\mathrm{Bol}/L_\mathrm{Edd}\sim1$--2 and
$\sim0.1$--0.2, respectively. Both cases under-predicted the Sw2/UVW2
flux by a factor of $\sim$20.

We then considered soft Comptonization to describe the spectra (no
hard Comptonization because of no clear hard X-ray emission). We found
that the data could not constrain all paramenters well. We thus only
report two special cases in Table~\ref{tbl:sppds107102}: the first
case assumes $M_\mathrm{BH}=10^5$ \msun\ (Model (e)), and the second case
includes the Sw2/UVW2 detection in the fits (Model (f)). Because the
optical depth $\tau$ of the corona was found to be consistent within
the 90\% errors, we tied it to be the same for all observations in
both cases. We found that the former case under-predicted the Sw2/UVW2
flux by a factor of 11. In the latter case, $M_\mathrm{BH}\sim
2\times10^6$ \msun\ and $L_\mathrm{Bol}/L_\mathrm{Edd}\sim0.09$--0.14 were
inferred.

\subsection{The \textit{Swift} Follow-up Observations}
\label{sec:swobs}
The source was not detected in the XRT in either \textit{Swift}
observation. We estimated a 0.3--2 keV count rate of
$(2^{+4})\times10^{-4}$ counts~s$^{-1}$ and $(5^{+8})\times10^{-4}$
counts~s$^{-1}$ (the errors are the 90\% confidence bounds) from Sw1
and Sw2, corresponding to a 0.3--10 keV absorbed luminosity of
($3^{+5}$)$\times$10$^{41}$ \ergsec\ and ($7^{+11}$)$\times$10$^{41}$
\ergsec\ (based on the MCD fit to XMM1), respectively. Thus the source
probably has become fainter. In the UV, we obtained the Sw1/UVW1 flux
of $(1.4\pm0.2)\times10^{-17}$ \ergscma\ (AB Mag of $22.7\pm0.2$), and
the Sw2/UVW2 flux of $(1.1\pm0.3)\times10^{-17}$ \ergscma\ (AB Mag of
$23.4\pm0.3$).

\section{DISCUSSION AND CONCLUSIONS}
\label{sec:conclusion}

We have shown that J1231+1106 exhibited a $\sim$3.8 hr QPO in the two
\xmm\ observations in 2005 at a $\sim$$5\sigma$ significance level,
making it one of the very few SMBHs with an X-ray QPO significantly
detected. Its rms generally increases with energy, a trend also
observed in most QPOs in Galactic BHBs and that in \object{RE
  J1034+396} \citep{remc2006,midowa2009}. Galactic BHBs can show both
low-frequency ($\sim$0.1--30 Hz) and high-frequency (40--450 Hz) QPOs
\citep{remc2006}. The former can have rms $>$15\%, while the latter
have rms $\sim$1\%. If the QPO in J1231+1106 corresponds to the
low-frequency type in BHBs, we infer $M_\mathrm{BH}<4\times10^6$
\msun. High-frequency QPOs in BHBs sometimes display in pair with
frequencies scaling in a 3:2 ratio and possibly following the relation
of $f_0=931(M_\mathrm{BH}/M_{\odot})^{-1}$ Hz, where $f_0$ is the
fundamental frequency of the pair \citep{remc2006}. Using this
relation, $M_\mathrm{BH}$ in J1231+1106 is $2.6\times10^7$ or
$3.8\times10^7$ \msun, depending on whether we observed the
periodicity of $2f_0$ or $3f_0$, respectively. If we assume instead
the QPO centroid frequency to be the Keplerian frequency at the ISCO,
$M_\mathrm{BH}$ would be within $3\times10^7$--$2\times10^8$ \msun,
depending on the spin of the BH. Considering the small BH mass
($\sim$$10^5$ \msun) inferred by \citet{hokite2012} from narrow
optical emission lines and the large rms, the QPO in J1231+1106 is
probably a low-frequency type. This QPO should have a different origin
from those in \object{RE J1034+396} and \object{Swift
  J164449.3+573451} because the latter two QPOs are contributed mostly
by the hard X-ray component \citep{miutdo2011,remire2012}. Considering
the probably high Eddington ratios of J1231+1106, the QPO could be due
to thermal instability, which was often used to explain the
``limit-cycle'' behavior of the BHB \object{GRS 1915+105} at high
accretion rates \citep[e.g.,][]{zhyugu2011}.

The ultrasoft X-ray spectra of J1231+1106 in the \xmm\ observations
appear broader than a single-temperature BB, but can be described with
either pure thermal disk emission or soft Comptonization, with $M_{\rm
  BH}$ consistent with that inferred from narrow optical emission
lines. The former model is supported by the observation of the disk
luminosity approximately following the relation of $L\propto T_{\rm
  MCD}^4$ expected for a standard disk, although the dynamical range
is small (only a factor of $\sim$2). The latter model is supported by
the detection of the QPO, which, in the case of Galactic BHBs, often
occurs in states with strong Comptonization \citep{remc2006}. Both
models under-predict the Sw2/UVW2 flux by more than one order of
magnitude, unless, for the soft Comptonization model, we assumed a BH
mass ($\sim$$10^6$ \msun) much larger than that inferred from narrow
optical emission lines and Eddington ratios ($\sim$0.1) too low to see
strong soft Comptonization \citep{tekaaw2012}. Thus the Sw2/UVW2 flux
is probably dominated by galaxy emission.

Therefore, both the strong fast variability, whose power is mostly in
the QPO, and the spectral modeling suggest J1231+1106 as a relatively
small BH accreting at high rates. This is a popular explanation for
narrow line Seyfert 1 galaxies \citep[NLS1s,
  e.g.,][]{bobrfi1996,grkole2010}, and J1231+1106 could be an extreme
case of such object, with only soft excess \citep[see
  also][]{tekaaw2012}. Alternatively, the source could be in a pure
thermal state hardly observed in AGNs. This scenario was suggested by
\citet{misaro2013} for \object{GSN 069}, a source showing many
similarities as J1231+1106 (e.g., no hard X-ray detection, strong fast
variability, and no broad H$_\alpha$ or H$_\beta$ lines) and having
UV-to-X-ray spectra consistent with pure thermal disk emission. In
either case, the non-detection of broad H$_\alpha$ or H$_\beta$ lines
in J1231+1106 should indicate their absence, instead of being
hiddened. This could be because the source had too weak hard X-ray
emission to maintain the broad line region in equilibrium, or because
the source only entered the current bright state recently, leaving no
enough time to form the mature broad line region \citep{misaro2013}.

However, ultrasoft X-ray spectra are more commonly seen in TDEs. Such
events are expected to rise on timescales of months and decay on
timescales of years, with the accretion rate approximately following
$t^{-5/3}$, where $t$ is the time since disruption, if the disrupted
stars are solar-type \citep{re1988,re1990}. About a dozen TDE
candidates have been found \citep[e.g.,][]{ko2002,licagr2011}. For
J1231+1106, XMM1 is $\sim$2.5 yr before XMM2 and XMM3, with comparable
fluxes. If it is a TDE, XMM1 should be in the rising phase, and XMM2
and XMM3 in the decay. The bolometric luminosity in 2013 would then
decrease from those in XMM2 and XMM3 by a factor of $\sim$10,
consistent with the \textit{Swift} observations. The absence of broad
H$_\alpha$ or H$_\beta$ lines in the Magellan spectrum in 2012 could
be because the source is too faint now, and the narrow emission lines
are light echo from the TDE. Alternatively, the disrupted star is an
evolved one, in which case the events have much longer timescales than
those involving solar-type stars \citep{magura2012} (this might also
explain \object{GSN 069}, which showed no significant variability in
the $\sim$1 yr monitoring by \textit{Swift} in 2010--2011, but at fluxes
a factor of $>$240 of those in the \textit{ROSAT} observations in
1994). Under the TDE explanation, the QPO could be due to the special
accretion mode in such events.

\acknowledgments Acknowledgments: We want to thank the \textit{Swift}
PI Neil Gehrels for approving our ToO request to observe J1231+1106
twice. This work is supported by NASA Grant NNX10AE15G.

\end{document}